# TCP Selective Acknowledgments and UBR Drop Policies to Improve ATM-UBR Performance over Terrestrial and Satellite Networks [*]

Rohit Goyal, Raj Jain, Shivkumar Kalyanaraman,
Sonia Fahmy, Bobby Vandalore, Sastri Kota[†]
Department of Computer Information Science
The Ohio State University
2015 Neil Avenue, DL 395
Columbus, OH 43210
E-mail: {*goyal,jain*}@cis.ohio-state.edu

## Abstract

*We study the performance of Selective Acknowledgments with TCP over the ATM-UBR service category. We examine various UBR drop policies, TCP mechanisms and network configurations to recommend optimal parameters for TCP over UBR. We discuss various TCP congestion control mechanisms compare their performance for LAN and WAN networks. We describe the effect of satellite delays on TCP performance over UBR and present simulation results for LAN, WAN and satellite networks. SACK TCP improves the performance of TCP over UBR, especially for large delay networks. Intelligent drop policies at the switches are an important factor for good performance in local area networks.*

## 1 Introduction

The Unspecified Bit Rate (UBR) service in ATM networks does not have any explicit congestion control mechanisms [2]. In the simplest form of UBR, switches drop cells whenever their buffers overflow. As a result, TCP connections using ATM-UBR service with limited switch buffers experience low throughput [3, 4, 5, 9, 13]. In our previous paper [9] we analyzed several enhancements to the UBR drop policies, and showed that these enhancements can improve the performance of TCP over UBR. We also analyzed the performance of Reno TCP (TCP with fast retransmit and recovery) over UBR, and concluded that fast retransmit and recovery hurts the performance of TCP in the presence of congestion losses over wide area networks.

This paper discusses the performance of TCP with selective acknowledgments (SACK TCP) over the UBR service category. We compare the performance of SACK TCP with vanilla TCP (TCP with slow start) and Reno TCP (TCP with slow start and fast retransmit and recovery). Simulation results of the performance the SACK TCP with several UBR drop policies over terrestrial and satellite links are presented.

Section 2 describes the TCP congestion control mechanisms including the Selective Acknowledgments (SACK) option for TCP. Section 3 describes our implementation of SACK TCP and Section 4 analyzes the features and retransmission properties of SACK TCP. We also describe a change to TCP's fast retransmit and recovery, proposed in [18, 22] and named "New Reno" in [18]. Section 7 discusses some issues relevant to the performance of TCP over satellite networks. The remainder of the paper presents simulation results comparing the performance of various TCP congestion avoidance methods.

## 2 TCP Congestion Control

TCP's congestion control mechanisms are described in detail in [15, 21]. TCP uses a window based flow control policy. The variable RCVWND is used as a measure of the receiver's buffer capacity. When a destination TCP host receives a segment, it sends an acknowledgment (ACK) for the next expected segment. TCP congestion control is built on this window based flow control. The following subsections describe the various TCP congestion control policies.

### 2.1 Slow Start and Congestion Avoidance

The sender TCP maintains a variable called congestion window (CWND) to measure the network capacity. The number of unacknowledged packets in the network is limited to CWND or RCVWND whichever is lower. Initially, CWND is set to one segment and it increases by one segment on the receipt of each new ACK until it reaches a maximum (typically 65536 bytes). It

---

can be shown that in this way, CWND doubles every round trip time, and this corresponds to an exponential increase in the CWND every round trip time [15].

The sender maintains a retransmission timeout for the last unacknowledged packet. Congestion is indicated by the expiration of the retransmission timeout. When the timer expires, the sender saves half the CWND in a variable called SSTHRESH, and sets CWND to 1 segment. The sender then retransmits segments starting from the lost segment. CWND is increased by one segment on the receipt of each new ACK until it reaches SSTHRESH. This is called the slow start phase. After that, CWND increases by one segment every round trip time. This results in a linear increase of CWND every round trip time, and is called the congestion avoidance phase. Figure 1 shows the slow start and congestion avoidance phases for at typical TCP connection.

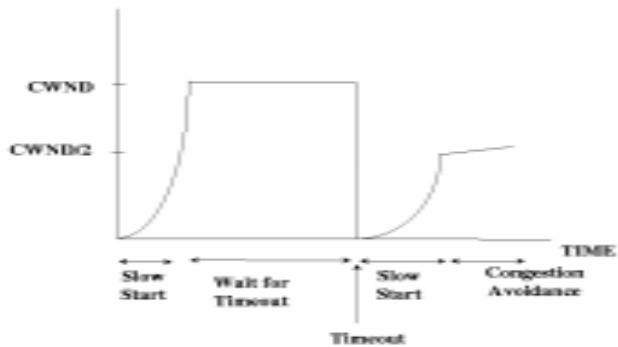

Figure 1: TCP Slow Start and Congestion Avoidance

## 2.2 Fast Retransmit and Recovery

Current TCP implementations use a coarse granularity (typically 500 ms) timer for the retransmission timeout. As a result, during congestion, the TCP connection can lose much time waiting for the timeout. In Figure 1, the horizontal CWND line shows the time lost in waiting for a timeout to occur. During this time, the TCP neither sends new packets nor retransmits lost packets. Moreover, once the timeout occurs, the CWND is set to 1 segment, and the connection takes several round trips to efficiently utilize the network. TCP Reno implements the fast retransmit and recovery algorithms that enable the connection to quickly recover from isolated segment losses [21].

If a segment is dropped by the network, the subsequent segments that arrive at the receiver are out-of-order segments. For each out-of-order segment, the TCP receiver immediately sends and ACK to the sender indicating the sequence number of the missing segment. This ACK is called a duplicate ACK. When the sender receives three duplicate ACKs, it concludes that the segment indicated by the ACKs has been lost, and immediately retransmits the lost segment. The sender then reduces CWND to half (plus 3 segments) and also saves half the original CWND value in SSTHRESH. Now for each subsequent duplicate ACK, the sender inflates CWND by one and tries to send a new segment. Effectively, the sender waits for half a round trip before sending one segment for each subsequent duplicate ACK it receives. As a result, the sender maintains the network pipe at half of its capacity at the time of fast retransmit.

Approximately one round trip after the missing segment is retransmitted, its ACK is received (assuming the retransmitted segment was not lost). At this time, instead of setting CWND to one segment and proceeding to do slow start until CWND reaches SSTHRESH, the TCP sets CWND to SSTHRESH, and then does congestion avoidance. This is called the fast recovery algorithm.

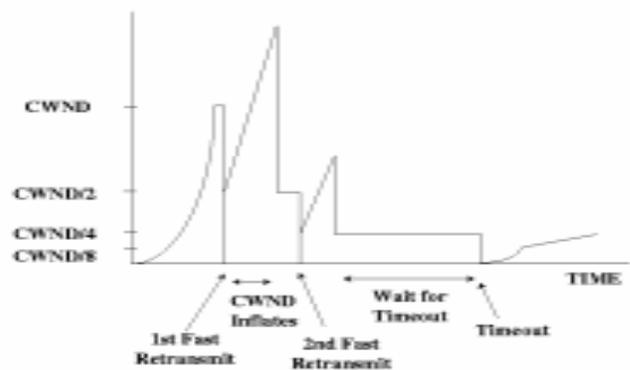

Figure 2: TCP Fast Retransmit and Recovery

## 2.3 A Modification to Fast Retransmit and Recovery: TCP New Reno

It has been known that fast retransmit and recovery cannot recover from multiple packet losses. Figure 2 shows a case when three consecutive packets are lost from a window, the sender TCP incurs fast retransmit twice and then times out. At that time, SSTHRESH is set to one-eighth of the original congestion window value (CWND in the figure). As a result, the exponential phase lasts a very short time, and the linear increase begins at a very small window. Thus, the TCP sends at a very low rate and loses much throughput.

The "fast-retransmit phase" was introduced in [22], in which the sender remembers the highest sequence number sent (RECOVER) when the fast retransmit was first triggered. After the first unacknowledged packet is retransmitted, the sender follows the usual fast recovery algorithm and inflates the CWND by one for each duplicate ACK it receives. When the sender receives an acknowledgment for the retransmitted packet, it checks if the ACK acknowledges all seg-ments including RECOVER. If so, the ACK is a new ACK, and the sender exits the fast retransmit-recovery phase, sets its CWND to SSTHRESH and starts a linear increase. If on the other hand, the ACK is a partial ACK, i.e., it acknowledges the retransmitted segment, and only a part of the segments before RECOVER, then the sender immediately retransmits the next expected segment as indicated by the ACK. This continues until all segments including RECOVER are acknowledged. **This mechanism ensures that the sender will recover from N segment losses in N round trips.**

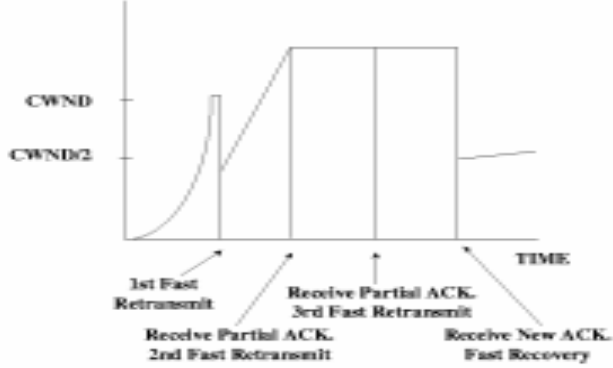

Figure 3: TCP with the fast retransmit phase

As a result, the sender can recover from multiple packet losses without having to timeout. In case of small propagation delays, and coarse timer granularities, this mechanism can effectively improve TCP throughput over vanilla TCP. Figure 3 shows the congestion window graph of a TCP connection for three contiguous segment losses. The TCP retransmits one segment every round trip time (shown by the CWND going down to 1 segment) until a new ACK is received.

### 2.4 Selective Acknowledgments

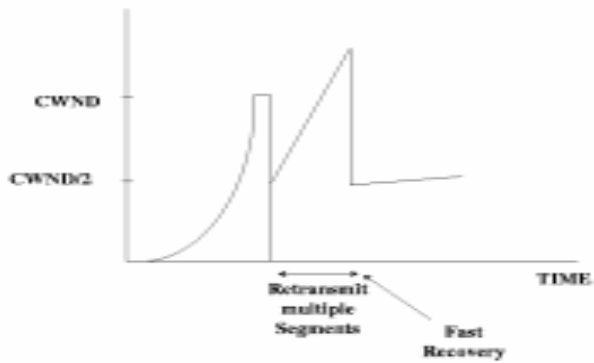

Figure 4: SACK TCP Recovery from packet loss

TCP with Selective Acknowledgments (SACK TCP) has been proposed to efficiently recover from multiple segment losses [20]. In SACK TCP, acknowledgments contain additional information about the segments that have been received by the destination. When the destination receives out-of-order segments, it sends duplicate ACKs (SACKs) acknowledging the out-of-order segments it has received. From these SACKs, the sending TCP can reconstruct information about the segments not received at the destination. When the sender receives three duplicate ACKs, it retransmits the first lost segment, and inflates its CWND by one for each duplicate ACK it receives. This behavior is the same as Reno TCP. However, when the sender is allowed to send a segment, it uses the SACK information to retransmit lost segments before sending new segments. As a result, the sender can recover from multiple dropped segments in about one round trip. Figure 4 shows the congestion window graph of a SACK TCP recovering from segment losses. During the time when the congestion window is inflating (after fast retransmit has incurred), the TCP is sending missing packets before any new packets.

## 3 SACK TCP Implementation

In this subsection, we describe our implementation of SACK TCP and some properties of SACK. Our implementation is based on the SACK implementation described in [18, 19, 20].

The SACK option is negotiated in the SYN segments during TCP connection establishment. The SACK information is sent with an ACK by the data receiver to the data sender to inform the sender of out-of-sequence segments received. The format of the SACK packet is described in [20]. The SACK option is sent whenever out of sequence data is received. All duplicate ACK's contain the SACK option. The option contains a list of some of the contiguous blocks of data already received by the receiver. Each data block is identified by the sequence number of the first byte in the block (the left edge of the block), and the sequence number of the byte immediately after the last byte of the block. Because of the limit on the maximum TCP header size, at most three SACK blocks can be specified in one SACK packet.

The receiver keeps track of all the out-of-sequence data blocks received. When the receiver generates a SACK, the first SACK block specifies the block of data formed by the most recently received data segment. This ensures that the receiver provides the most up-to-date information to the sender. After the first SACK block, the remaining blocks can be filled in any order.

The sender also keeps a table of all the segments sent but not ACKed. When a segment is sent, it is entered into the table. When the sender receives an ACK with the SACK option, it marks in the table all the segments specified in the SACK option blocks as SACKed. The entries for each segment remain in the table until the segment is ACKed. The remaining behavior of the sender is very similar to Reno implementations with the modification suggested in Section 2.3 [1]. When the sender receives three duplicate ACKs, it retransmits the first unacknowledged packet. During the fast retransmit phase, when the sender is sending one segment for each duplicate ACK received, it first tries to retransmit the holes in the SACK blocks before sending any new segments. When the sender retransmits a segment, it marks the segment as retransmitted in the table. If a retransmitted segment is lost, the sender times out and performs slow start. When a timeout occurs, the sender resets the table.

During the fast retransmit phase, the sender maintains a variable PIPE that indicates how many bytes are currently in the network pipe. When the third duplicate ACK is received, PIPE is set to the value of CWND and CWND is reduced by half. For every subsequent duplicate ACK received, PIPE is decremented by one segment because the ACK denotes a packet leaving the pipe. The sender sends data (new or retransmitted) only when PIPE is less than CWND. This implementation is equivalent to inflating the CWND by one segment for every duplicate ACK and sending segments if the number of unacknowledged bytes is less than the congestion window value.

When a segment is sent, PIPE is incremented by one. When a partial ACK is received, PIPE is decremented by two. The first decrement is because the partial ACK represents a retransmitted segment leaving the pipe. The second decrement is done because the original segment that was lost, and had not been accounted for, is now actually considered to be lost.

## 4 TCP: Analysis of Recovery Behavior

In this section, we discuss the behavior of SACK TCP. We first analyze the properties of Reno TCP and then lead into the discussion of SACK TCP. Vanilla TCP without fast retransmit and recovery (we refer to TCP with only slow start and congestion avoidance as vanilla TCP), will be used as the basis for comparison. Every time congestion occurs, TCP tries to reduce its CWND window by half and then enters congestion avoidance. In the case of vanilla TCP, when a segment is lost, a timeout occurs, and the congestion window reduces to one segment. From there, it takes about $log_2(\text{CWND}/(2 \times \text{TCP segment size}))$ round trip times (RTTs) for CWND to reach the target value. This behavior is unaffected by the number of segments lost from a particular window.

---

[1] It is not clear to us whether the SACK option provides better performance with or without New Reno. This is under further study.

### 4.1 Reno TCP

When a single segment is lost from a window, Reno TCP recovers within approximately one RTT of knowing about the loss or two RTTs after the lost packet was first sent. The sender receives three duplicate ACKs about one RTT after the dropped packet was sent. It then retransmits the lost packet. For the next round trip, the sender receives duplicate ACKs for the whole window of packets sent after the lost packet. The sender waits for half the window and then transmits a half window worth of new packets. All of this takes about one RTT after which the sender receives a new ACK acknowledging the retransmitted packet and the entire window sent before the retransmission. CWND is set to half its original value and congestion avoidance is performed.

When multiple packets are dropped, Reno TCP cannot recover and may result in a timeout. The fast retransmit phase modification can recover from multiple packet losses by retransmitting a single packet every round trip time.

### 4.2 SACK TCP

In this subsection we show that SACK TCP can recover from multiple packet losses more efficiently than Reno or vanilla TCP.

Suppose that at the instant when the sender learns of the first packet loss (from three duplicate ACKs), the value of the congestion window is CWND. Thus, the sender has CWND bytes of data waiting to be acknowledged. Suppose also that the network drops a block of data which is CWND/n bytes long (This will typically result in several segments being lost). After one RTT of sending the first dropped segment, the sender receives three duplicate ACKs for this segment. It retransmits the segment, sets PIPE to CWND − 3, and sets CWND to CWND/2. For each duplicate ACK received, PIPE is decremented by 1. When PIPE reaches CWND, then for each subsequent duplicate ACK received, another segment can be sent. All the ACKs from the previous window take 1 RTT to return. For one half RTT nothing is sent (since PIPE > CWND). For the next half RTT, if CWND/n bytes were dropped, then only CWND/2 − CWND/n bytes (of retransmitted or new segments) can be sent. Thus, all the dropped segments can be retransmitted in 1 RTT if

$$\text{CWND}/2 - \text{CWND}/n \geq \text{CWND}/n$$

i.e., $n \geq 4$. Therefore, for SACK TCP to be able to retransmit all lost segments in one RTT, the network can drop at most CWND/4 bytes from a window of CWND.

Now, we calculate the maximum amount of data that can be dropped for SACK TCP to be able to retransmit everything in two RTTs. Suppose again that CWND/n bytes are dropped from a window of size CWND. Then, in the first RTT from receiving the 3 duplicate ACKs, the sender can retransmit upto CWND/2 − CWND/n bytes. In the second RTT, the sender can retransmit 2(CWND/2 − CWND/n) bytes. This is because for each retransmitted segment in the first RTT, the sender receives a partial ACK that indicates that the next segment is missing. As a result, PIPE is decremented by 2, and the sender can send 2 more segments (both of which could be retransmitted segments) for each partial ACK it receives. Thus, all the dropped segments can be retransmitted in 2 RTTs if

$$\frac{\text{CWND}}{2} - \frac{\text{CWND}}{n} + 2(\frac{\text{CWND}}{2} - \frac{\text{CWND}}{n}) \geq \frac{\text{CWND}}{n}$$

i.e. $n \geq 8/3$. This means that at most $3\times$CWND/8 bytes can be dropped from a window of size CWND for SACK TCP to be able to recover in 2 RTTs.

Generalizing the above argument, we have the following result: **The number of RTTs needed by SACK TCP to recover from a loss of CWND/n is at most $\lceil \log (n/(n-2)) \rceil$ for n > 2**. If more than half the CWND is dropped, then there will not be enough duplicate ACKs for PIPE to become large enough to transmit any segments in the first RTT. Only the first dropped segment will be retransmitted on the receipt of the third duplicate ACK. In the second RTT, the ACK for the retransmitted packet will be received. This is a partial ACK and will result in PIPE being decremented by 2 so that 2 packets can be sent. As a result, PIPE will double every RTT, and **SACK will recover no slower than slow start** [**18, 19**]. SACK would still be advantageous because timeout would be still avoided unless a retransmitted packet were dropped.

## 5 The ATM-UBR Service

The basic UBR service can be enhanced by implementing intelligent drop policies at the switches. A comparative analysis of various drop policies on the performance of Vanilla and Reno TCP over UBR is presented in [9]. Section 5.3 briefly summarizes the results of our earlier work. This section briefly describes the drop policies.

### 5.1 Early Packet Discard

The Early Packet Discard policy [1] maintains a threshold R, in the switch buffer. When the buffer occupancy exceeds R, then all new incoming packets are dropped. Partially received packets are accepted if possible. It has been shown [9] that EPD improves the efficiency of TCP over UBR but does not improve fairness. The effect of EPD is less pronounced for large delay-bandwidth networks. In satellite networks, EPD has little or no effect in the performance of TCP over UBR.

### 5.2 Selective Packet Drop and Fair Buffer Allocation

These schemes use per-VC accounting to maintain the current buffer utilization of each UBR VC. A fair allocation is calculated for each VC, and if the VC's buffer occupancy exceeds its fair allocation, its subsequent incoming packet is dropped. Both schemes maintain a threshold R, as a fraction of the buffer capacity K. When the total buffer occupancy exceeds R×K, new packets are dropped depending on the $VC_i$'s buffer occupancy ($Y_i$). In the Selective Drop scheme, a VC's *entire packet* is dropped if

$$(X > R) \text{ AND } (Y_i \times N_a/X > Z)$$

where $N_a$ is the number of active VCs (VCs with at least one cell the buffer), and Z is another threshold parameter ($0 < Z \leq 1$) used to scale the effective drop threshold.

The Fair Buffer Allocation proposed in [8] is similar to Selective Drop and uses the following formula:

$$(X > R) \text{ AND } (Y_i \times N_a/X > Z \times ((K - R)/(X - R)))$$

### 5.3 Performance of TCP over UBR: Summary of Earlier Results

In our earlier work [9, 10] we discussed the following results:

- For multiple TCP connections, the switch requires a buffer size of the sum of the receiver windows of the TCP connections.
- With limited buffers, TCP over plain UBR results in poor performance.
- TCP performance over UBR can be improved by intelligent drop policies like Early Packet Discard, Selective Drop and Fair Buffer Allocation.
- TCP fast retransmit and recovery improves TCP performance over LANs, and actually degrades performance over WANs in the presence of congestion losses.

## 6 Simulation Results with SACK TCP over UBR

This section presents the simulation results of the various enhancements of TCP and UBR presented in the previous sections.

### 6.1 The Simulation Model

All simulations use the N source configuration shown in Figure 5. All sources are identical and persistent TCP sources i.e., the sources always send a segment as long as it is permitted by the TCP window. Moreover, traffic is unidirectional so that only the sources send data. The destinations only send ACKs. The performance of TCP over UBR with bidirectional traffic is a topic of further study. The delayed acknowledgment

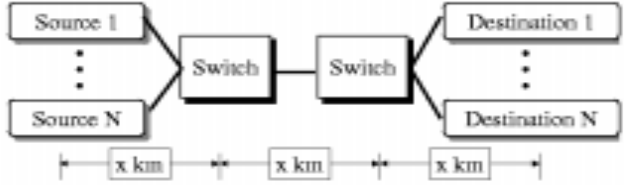

Figure 5: The N source TCP configuration

timer is deactivated, i.e., the receiver sends an ACK as soon as it receives a segment.

Link delays are 5 microseconds for LAN configurations and 5 milliseconds for WAN configurations. This results in a round trip propagation delay of 30 microseconds for LANs and 30 milliseconds for WANs respectively. The TCP segment size is set to 512 bytes. For the LAN configurations, the TCP maximum window size is limited by a receiver window of 64K bytes. This is the default value specified for TCP implementations. For WAN configurations, a window of 64K bytes is not sufficient to achieve 100% utilization. We therefore use the window scaling option to specify a maximum window size of 600,000 Bytes. This window is sufficient to provide full utilization with each TCP source.

All link bandwidths are 155.52 Mbps, and Peak Cell Rate at the ATM layer is 155.52 Mbps. The duration of the simulation is 10 seconds for LANs and 20 seconds for WANs. This allows enough round trips for the simulation to give stable results.

The configurations for satellite networks are discussed in Section 7.

### 6.2 Performance Metrics

The performance of the simulation is measured at the TCP layer by the Efficiency and Fairness as defined below.

$$\text{Efficiency} = \frac{\text{(Sum of TCP throughputs)}}{\text{(Maximum possible TCP throughput)}}$$

TCP throughput is measured at the destination TCP layer as the total number of bytes delivered to the application divided by the simulation time. This is divided by the maximum possible throughput attainable by TCP. With 512 bytes of TCP data in each segment, 20 bytes of TCP header, 20 bytes of IP header, 8 bytes of LLC header, and 8 bytes of AAL5 trailer are added. This results in a net possible throughput of 80.5% of the ATM layer data rate or 125.2 Mbps on a 155.52 Mbps link.

$$\text{Fairness Index} = (\Sigma x_i)^2 / (\text{N} \times \Sigma x_i^2)$$

Where $x_i$ is the ratio of the achieved throughput to the expected throughput of the $i$th TCP source, and N is the number of TCP sources. Fairness values close to 0.99 indicate near perfect fairness.

### 6.3 Simulation Results

We performed simulations for the LAN and WAN configurations for three drop policies – vanilla UBR (switches drop incoming cells when their buffers overflow), Early Packet Discard (EPD) and Selective Drop. For LANs, we used buffer sizes of 1000 and 3000 cells. These are representative of the typical buffer sizes in current switches. For WANs, we chose buffer sizes of approximately one and three times the bandwidth – round trip delay product. Tables 1 and 2 show the efficiency and fairness values of SACK TCP with various UBR drop policies. Several observations can be made from these tables:

- For most cases, for a given drop policy, **SACK TCP provides higher efficiency than either the corresponding drop policy in vanilla or Reno TCP.** This confirms the intuition provided by the analysis of SACK that SACK recovers at least as fast as slow start when multiple packets are lost. In fact, for most cases, SACK recovers faster than both fast retransmit/recovery and slow start algorithms.

- **For LANs, the effect of drop policies is very important and can dominate the effect of SACK**. For UBR with tail drop, SACK provides a significant improvement over Vanilla and Reno TCPs. However, as the drop policies get more sophisticated, the effect of TCP congestion mechanism is less pronounced. This is because, the typical LAN switch buffer sizes are small compared to the default TCP maximum window of 64K bytes, and so buffer management becomes a very important factor. Moreover, the degraded performance of SACK in few cases can be attributed to excessive timeout due to the retransmitted packets being lost. In this case SACK loses several round trips in retransmitting parts of the lost data and then times out. After timeout, much of the data is transmitted again, and this results in wasted throughput. This result reinforces the need for a good drop policy for TCP over UBR.

- **The throughput improvement provided by SACK is more significant for wide area networks**. When propagation delay is large, a timeout results in the loss of a significant amount of time during slow start from a window of one segment. With Reno TCP (with fast retransmit and recovery), performance is further degraded (for multiple packet losses) because timeout occurs at a much lower window than vanilla TCP. With SACK TCP, a timeout is avoided at many times, and recovery is complete within a short number of roundtrips. Even if timeout occurs, the recovery

Table 1: SACK TCP over UBR : Efficiency

| Config-uration | Num of Srcs | Buffer (cells) | UBR | EPD | Sel Drop |
|---|---|---|---|---|---|
| LAN | 5 | 1000 | 0.76 | 0.85 | 0.94 |
| LAN | 5 | 3000 | 0.98 | 0.97 | 0.98 |
| LAN | 15 | 1000 | 0.57 | 0.78 | 0.91 |
| LAN | 15 | 3000 | 0.86 | 0.94 | 0.97 |
| **SACK Column Average** | | | 0.79 | 0.89 | 0.95 |
| **Vanilla TCP Average** | | | 0.34 | 0.67 | 0.84 |
| **Reno TCP Average** | | | 0.69 | 0.97 | 0.97 |
| WAN | 5 | 12,000 | 0.90 | 0.88 | 0.95 |
| WAN | 5 | 36,000 | 0.97 | 0.99 | 1.00 |
| WAN | 15 | 12,000 | 0.93 | 0.80 | 0.88 |
| WAN | 15 | 36,000 | 0.95 | 0.95 | 0.98 |
| **SACK Column Average** | | | 0.94 | 0.91 | 0.95 |
| **Vanilla TCP Average** | | | 0.91 | 0.9 | 0.91 |
| **Reno TCP Average** | | | 0.78 | 0.86 | 0.81 |

Table 2: SACK TCP over UBR : Fairness

| Config-uration | Num of Srcs | Buffer (cells) | UBR | EPD | Sel Drop |
|---|---|---|---|---|---|
| LAN | 5 | 1000 | 0.22 | 0.88 | 0.98 |
| LAN | 5 | 3000 | 0.92 | 0.97 | 0.96 |
| LAN | 15 | 1000 | 0.29 | 0.63 | 0.95 |
| LAN | 15 | 3000 | 0.74 | 0.88 | 0.98 |
| **SACK Column Average** | | | 0.54 | 0.84 | 0.97 |
| **Vanilla TCP Average** | | | 0.69 | 0.69 | 0.92 |
| **Reno TCP Average** | | | 0.71 | 0.98 | 0.99 |
| WAN | 5 | 12,000 | 0.96 | 0.98 | 0.95 |
| WAN | 5 | 36,000 | 1.00 | 0.94 | 0.99 |
| WAN | 15 | 12,000 | 0.99 | 0.99 | 0.99 |
| WAN | 15 | 36,000 | 0.98 | 0.98 | 0.96 |
| **Column Average** | | | 0.98 | 0.97 | 0.97 |
| **Vanilla TCP Average** | | | 0.76 | 0.95 | 0.94 |
| **Reno TCP Average** | | | 0.90 | 0.97 | 0.99 |

is as fast as slow start but a little time may be lost in the earlier retransmission.

- **The performance of SACK TCP can be improved by intelligent drop policies like EPD and Selective Drop**. This is consistent with our earlier results in [9]. Thus, we recommend that intelligent drop policies be used in UBR service.

- **The fairness values for selective drop are comparable to the values with the other TCP versions**. Thus, SACK TCP does not hurt the fairness in TCP connections with an intelligent drop policy like selective drop. The fairness of tail drop and EPD are sometimes a little lower for SACK TCP. This is again because retransmitted packets are lost and some connections timeout. Connections which do not timeout do not have to go through slow start, and thus can utilize more of the link capacity. The fairness among a set of hybrid TCP connections is a topic of further study.

## 7 Effects of Satellite Delays on TCP over UBR

Since TCP congestion control is inherently limited by the round trip time, long delay paths have significant effects on the performance of TCP over ATM. A large delay-bandwidth link must be utilized efficiently to be cost effective. This section discusses some of the issues that arise in the congestion control of large delay-bandwidth links. Simulation results of TCP over UBR with satellite delays are also presented. Related results in TCP performance over satellite are available in [23].

### 7.1 Window Scale Factor

The default TCP maximum window size is 65535 bytes. For a 155.52 Mbps ATM satellite link (with a propagation RTT of about 550 ms), a congestion window of about 8.7M bytes is needed to fill the whole pipe. As a result, the TCP window scale factor must be used to provide high link utilization. In our simulations, we use a receiver window of 34,000 and a window scale factor of 8 to achieve the desired window size.

### 7.2 Large Congestion Window and the congestion avoidance phase

During the congestion avoidance phase, CWND is incremented by 1 segment every RTT. Most TCP implementations follow the recommendations in [15], and increment by CWND by 1/CWND segments for each ACK received during the congestion avoidance. Since CWND is maintained in bytes, this increment translates to an increment of MSS×MSS/CWND bytes on the receipt of each new ACK. All operations are done on integers, and this expression avoids the need for floating point calculations. However, in the case of large delay-bandwidth paths where the window scale factor is used, MSS×MSS may be less than CWND. For example, with MSS = 512 bytes, MSS×MSS = 262144, and when CWND is larger than this value, the expression MSS×MSS/CWND yields zero. As a result, CWND is never increases during the congestion avoidance phase.

There are several solutions to this problem. The most intuitive is to use floating point calculations. This increases the processing overhead of the TCP layer and is thus undesirable. A second option is to not increment CWND for each ACK, but to wait for N ACKs such that N×MSS×MSS > CWND and then increment CWND by N×MSS×MSS/CWND. We call this the ACK counting option.

Another option would be to increase MSS to a larger value so that MSS×MSS would be larger than CWND at all times. The MSS size of the connection is limited by the smallest MTU of the connection. Most future TCPs are expected to use Path-MTU discovery to find out the largest possible MSS that can be used. This value of MSS may or may not be sufficient to ensure the correct functioning of congestion avoidance without ACK counting. Moreover, if TCP is running over a connectionless network layer like IP, the MTU may change during the lifetime of a connection and segments may be fragmented. In a cell based network like ATM, TCP could used arbitrary sized segments without worrying about fragmentation. The value of MSS can also have an effect on the TCP throughput, and larger MSS values can produce higher throughput. The effect of MSS on TCP over satellite is a topic of current research.

Table 3: TCP over UBR with Satellite Delays: Efficiency

| TCP | Num of Srcs | Buffer (cells) | UBR | EPD | Sel Drop |
|---|---|---|---|---|---|
| SACK | 5 | 200,000 | 0.86 | 0.6 | 0.72 |
| SACK | 5 | 600,000 | 0.99 | 1.00 | 1.00 |
| Reno | 5 | 200,000 | 0.84 | 0.12 | 0.12 |
| Reno | 5 | 600,000 | 0.30 | 0.19 | 0.22 |
| Vanilla | 5 | 200,000 | 0.70 | 0.73 | 0.73 |
| Vanilla | 5 | 600,000 | 0.88 | 0.81 | 0.82 |

Table 4: SACK TCP over UBR with Satellite Delays: Fairness

| Configuration | Num of Srcs | Buffer (cells) | UBR | EPD | Sel Drop |
|---|---|---|---|---|---|
| SACK | 5 | 200,000 | 1.00 | 0.83 | 0.94 |
| SACK | 5 | 600,000 | 1.00 | 1.00 | 1.00 |
| Reno | 5 | 200,000 | 0.96 | 0.97 | 0.97 |
| Reno | 5 | 600,000 | 1.00 | 1.00 | 1.00 |
| Vanilla | 5 | 200,000 | 1.00 | 0.87 | 0.89 |
| Vanilla | 5 | 600,000 | 1.00 | 1.00 | 1.00 |

## 8  Simulation Results of TCP over UBR in Satellite networks

The satellite simulation model is very similar to the model described in section 6.1. The differences are listed below:

- The link between the two switches in Figure 5 is now a satellite link with a one-way propagation delay of 275 ms. The links between the TCP sources and the switches are 1 km long. This results in a round trip propagation delay of about 550 ms.
- The maximum value of the TCP receiver window is now 8,704,000 bytes. This window size is sufficient to fill the 155.52 Mbps pipe.
- The TCP maximum segment size is 9180 bytes. A larger value is used because most TCP connections over ATM with satellite delays are expected to use larger segment sizes.
- The buffer sizes used in the switch are 200,000 cells and 600,000 cells. These buffer sizes reflect buffers of about 1 RTT and 3 RTTs respectively.
- The duration of simulation is 40 seconds.

Tables 3 and 4 show the efficiency and fairness values for Satellite TCP over UBR with 5 TCP sources and buffer sizes of 200,000 and 600,000 cells. Several observations can be made from the tables:

- **Selective acknowledgments significantly improve the performance of TCP over UBR for satellite networks**. The efficiency and fairness values are typically higher for SACK than for Reno and vanilla TCP. This is because SACK often prevents the need for a timeout and can recover quickly from multiple packet losses.
- **Fast retransmit and recovery is detrimental to the performance of TCP over large delay-bandwidth links**. The efficiency numbers for Reno TCP in table 3 are much lower than those of either SACK or Vanilla TCP. This reinforces the WAN results in table 1 for Reno TCP. Both the tables are also consistent with analysis in Figure 2, and show that fast retransmit and recovery cannot recover from multiple losses in the same window.
- Intelligent drop policies have little effect on the performance of TCP over UBR satellite networks. Again, these results are consistent with the WAN results in tables 1 and 2. The effect of intelligent drop policies is most significant in LANs, and the effect decreases in WANs and satellite networks. This is because LAN buffer sizes (1000 to 3000 cells) are much smaller compared to the default TCP maximum window size of 65535 bytes. For WANs and satellite networks, the switch buffer sizes and the TCP maximum congestion window sizes are both of the order of the round trip delays. As a result, efficient buffer management becomes more important for LANs than WANs and satellite networks.

## 9  Summary

This paper describes the performance of SACK TCP over the ATM UBR service category. SACK TCP is seen to improve the performance of TCP over UBR. UBR drop policies are also essential to improving the performance of TCP over UBR. As a result, TCP performance over UBR can be improved by either improving TCP using selective acknowledgments, or by introducing intelligent buffer management policies at the switches. Efficient buffer management has a more sig-nificant influence on LANs because of the limited buffer sizes in LAN switches compared to the TCP maximum window size. In WANs and satellite networks, the drop policies have a smaller impact because both the switch buffer sizes and the TCP windows are of the order of the bandwidth-delay product of the network. SACK TCP is especially helpful in satellite networks, and provides a large gain in performance over fast retransmit and recovery and slow start algorithms.

---

[2] All our papers and ATM Forum contributions are available from http://www.cis.ohio-state.edu/~jain